\documentclass[12pt]{article}
\usepackage{amsfonts}
\usepackage{pstricks}
\usepackage{pst-node}
\usepackage{epsfig}
\usepackage{amsmath,amssymb,latexsym}
\usepackage{xcolor}
\usepackage{hyperref}

\usepackage{import}

\usepackage{enumerate}

\usepackage{physics}

\usepackage[lmargin=1.0in, rmargin=1.0in, tmargin=1.0in, bmargin=1.0in]{geometry}

\usepackage{cancel}

\usepackage{empheq}

\usepackage{comment}

\usepackage{setspace}
\begin{document}



\title{\bf  The non-minimally coupled  symmetric teleparallel gravity with electromagnetic field}
\author{ Beyda Doyran, Özcan Sert, Muzaffer Adak  \\
  {\small $^1$Department of Physics, Faculty of Science, Pamukkale University, Denizli, Türkiye} \\
      {\small {\it E-mail:} {\blue beyda\_d@hotmail.com, osert@pau.edu.tr, madak@pau.edu.tr}}}

  \vskip 1cm
\date{\today}
\maketitle
\thispagestyle{empty}
\begin{abstract}
 \noindent
We construct  a symmetric teleparallel gravity  model which is non-minimally coupled with electromagnetic field  in four dimensions inspired by its Riemannian equivalent. We derive the field equations by taking the variation of this model, which is written here for the first time.   Then, we find some classes of spherically symmetric static solutions by  the coincident gauge of symmetric teleparallel spacetime.  \\



\end{abstract}

\section{Introduction}

General Theory of Relativity (GR), introduced by Einstein, has become a very popular theory with its success in explaining problems that Newton's theory could not solve and its compatibility with observational results in the solar system.
This theory was constructed in Riemannian geometries, where only curvature is different from zero, from the geometrical trio of curvature, torsion and non-metricity. In Riemannian geometry, the field equations of this theory can be derived by the Lagrangian formulation from an action known as the Einstein-Hilbert action.
However, there are still  big questions which are not found out by GR. Especially, in cosmology the accelerated expansion of the universe \cite{peebles-ratra-2003} and in astrophysics the presence of a mysterious form of matter which cannot be observed directly \cite{trimble-1987}, which are known as  dark energy and dark matter issues, are the most known among them. Meanwhile it is well known that geometric construction of gravity is not unique.
Even if GR survives today at some scales, to be able to describe sufficiently and  accurately  at all scales of  the universe, it may be challenged by alternative theories. Metric (or Weitzenböck) teleparallel, symmetric teleparallel and general teleparallel equivalent  of gravity can be constructed in non-Riemannian geometries by using only torsion, only non-metricity and both, respectively. One can consult for the literature \cite{adak-dereli-2023}. In order to remedy the above mentioned problems, among the non-Riemannian geometries, we choose to work with vanishing curvature and torsion, but non-vanishing non-metricity, known as  the symmetric teleparallel geometries.
The  gravity model in this geometry was first introduced as symmetric teleparallel gravity (STPG) in \cite{nester-1999}. This model was later constructed and its spherically symmetric solutions were founded  in the papers \cite{Adak2005,Adak2006_1,Adak20062}. 
This model has subsequently become a remarkable research topic, which has been extensively studied in many papers \cite{adak2008,adak2013ijmpa,tomi-lavinia2017,Jimenez2019,adak2018,koivisto-Iosifidis2018,koivisto2020,ditta-tiecheng-2023-epjc,caglar_ozcan_muzo_2022,adak-ozdemir-2023} and the references therein.

On the other hand, the investigation of such highly magnetic objects as magnetar, pulsar,  neutron star needs  a gravity model coupled  with Maxwell Lagrangian. Minimal  coupling  has  already been studied and it has the Reissner-Nordström solution. But non-minimal couplings offers a new research area and interesting solutions to the gravitoelectromagnetism problems. The detailed analysis of the non-minimally coupled electromagnetic field to gravity and  the corresponding  cosmological and astrophysical solutions to the model are studied in \cite{Dereli20111,Sert12Plus,Sert13MPLA,Dereli20112,Sert16regular,Sert2017} with Riemannian geometry.  After constructing the symmetric teleparallel  equivalent of GR, it is natural to ask the  investigation of the non-minimally coupled electromagnetic field to STPG. This  is the main motivation of this paper.
We have preferred to use the language of exterior algebra through the paper. After we give some basic notations and conventions of exterior algebra in the next section, 
we define the Lagrangian 4-form of the non-minimally coupled STPG and derive   field equations by variation. Then we search the spherically symmetric static solutions of  the non-minimal gravity model by using the coincident gauge of the symmetric teleparallel geometry.

\section{Non-Riemannian Geometry with Differential Forms }

In non-Riemannian geometry the space-time is described by the triple of differentiable manifold $M$,  symmetric and non-degenerate metric tensor $g$ and full (affine) connection $\omega^a{}_b$ 1-form that is metric independent \cite{thirring1997}, \cite{frankel2012}. Here the metric is expressed as $g=\eta_{ab} e^a \otimes e^b$ where $e^a$ is the $g$-orthonormal 1-form and $\eta_{ab}$ is the Minkowski metric having the signature $(-,+,+,+)$. 
In notation of exterior differential forms $d$, $D$, $\wedge$ symbols  represent  the exterior derivative, the Lorentz-covariant exterior derivative and the exterior product, respectively.
The Cartan structure equations are given by the following tensor-valued  curvature 2-form,  non-metricity 1-form, torsion 2-form, respectively 
\begin{subequations}\label{eq:cartan-ort}
 \begin{align}
     R^a{}_b &:= D\omega^a{}_b := d \omega^a{}_b + \omega^a{}_c \wedge \omega^c{}_b, \label{eq:curv}\\
     Q_{ab} &:= -\frac{1}{2} D\eta_{ab} = \frac{1}{2} (\omega_{ab} + \omega_{ba}), \label{eq:nonmetric}\\
     T^a &:= De^a = de^a + \omega^a{}_b \wedge e^b. \label{eq:tors}
 \end{align}
 \end{subequations}
By computing their covariant exterior derivative, one can obtain the Bianchi identities
  \begin{align}
      DQ_{ab}= \frac{1}{2}(R_{ab}+R_{ba}) , \qquad DT^a = R^a{}_b \wedge e^b , \qquad DR^a{}_b=0  \;.
  \end{align} 

The full connection  $\omega^a{}_b$  is separated to  a Riemannian part $\widetilde{\omega}_{ab}(g)$ depending on only metric and a non-Riemannian part $\mathrm{L}_{ab}(T,Q)$  depending on torsion and non-metricity as follow \cite{tucker1995},\cite{hehl1995},\cite{adak2022ijgmmp}
\begin{align}
     \omega_{ab}=\widetilde{\omega}_{ab} + \mathrm{L}_{ab}. \label{eq:connec-decom}
 \end{align}
Here  $\widetilde{\omega}_{ab} = -\widetilde{\omega}_{ba}$  is known as  the Levi-Civita connection 1-form and it can be obtained from orthonormal 1-forms \begin{equation}\label{eq:Levi-Civita}
   \widetilde{\omega}_{ab} = \frac{1}{2} \left[ -\iota_a de_b + \iota_b de_a + (\iota_a \iota_b de_c) e^c \right] \qquad 
 \end{equation}
and  the remainder of ${\omega}_{ab}$ is called the tensor-valued distortion 1-form 
 \begin{equation}
     \mathrm{L}_{ab} = \underbrace{  Q_{ab} + ( \imath_b Q_{ac} - \imath_a Q_{bc} ) e^c}_{disformation} + \underbrace{ \frac{1}{2} \left[ \iota_a T_b - \iota_b T_a - (\iota_a \iota_b T_c) e^c \right]}_{contortion}
 \end{equation}  
where $\iota_a$ denotes the interior product. The former component containing non-metricity is known disformation 1-form and the latter one containing torsion is contortion 1-form.

It is worthy to notice that the symmetric part of the affine connection  $\omega_{ab}$ is determined by only non-metricity, $\omega_{(ab)}= Q_{ab}$, the remainder part of $\omega_{ab}$ is the anti-symmetric. Here $\omega_{(ab)}:= \frac{1}{2} (\omega_{ab}+\omega_{ba}) = Q_{ab}$ is the reason why we  take the factor $-\frac{1}{2}$ in the definition of non-metricity (\ref{eq:nonmetric}).

\section{The Non-minimally Coupled  Symmetric Teleparallel Gravity with Electromagnetic Field}

Inspired by the symmetric teleparallel equivalent of general relativity \cite{Adak20062} we replace the Riemannian curvature scalar $R *1$ with  $\mathbb{Q}=-\mathcal{Q}*1 $.     
Then  similar to the non-minimal $Y(R)\mathbb{F}$ models 
\cite{Dereli20111,Sert12Plus,Sert13MPLA,Dereli20112,Sert16regular,Sert2017}, we consider  $Y(\mathcal{Q}) \mathbb{F}$ models in this paper,  where $\mathbb{F}$ is the Maxwell Lagrangian given in (\ref{Maxwell Lag}).   We also add  cosmological constant  $\Lambda$ and constrain torsion and curvature to zero by $\lambda_a$ and $\rho^b{}_a$ Lagrange multiplier 2-forms. Therefore, we start  with the following Lagrangian 4-form of the  non-minimal model 
\begin{align} \label{eq:lag-stpg}
      L= \mathbb{Q} - Y(\mathcal{Q}) \mathbb{F} + \Lambda *1 + T^a \wedge \lambda_a + R^a{}_b \wedge \rho^b{}_a .
 \end{align}
Here  $\mathbb{Q}$ is expressed explicitly by the STPG Lagrangian 4-form 
 \begin{align} \label{eq:lag-QQ}
    \mathbb{Q} =& \, c_1  Q_{ab} \wedge *Q^{ab} + c_2 \left(Q_{ab} \wedge e^b\right) \wedge *\left(Q^{ac} \wedge e_c\right) + c_3 \left(Q_{ab} \wedge e_c \right) \wedge *\left(Q^{ac} \wedge e^b\right)  \nonumber \\
     & \quad + c_4 Q \wedge *Q   + c_5 (Q \wedge e^b) \wedge * (Q_{ab} \wedge e^a) 
 \end{align}
where $c_i$, $i=1,\cdots,5$, are coupling constants and $Q=\eta_{ab}Q^{ab}=\eta_{ab}\omega^{ab}$ is the non-metricity trace 1-form. Its Hodge dual yields a zero-form $\mathcal{Q}$, i.e., $\mathcal{Q}=* \mathbb{Q}$ or $\mathbb{Q}= - \mathcal{Q}*1$. Besides, $\mathbb{F}$ is the Maxwell Lagrangian 4-form
  \begin{align}\label{Maxwell Lag}
      \mathbb{F} = F \wedge *F
  \end{align}
where $F$ denotes the Maxwell electromagnetic 2-form which is written in terms of potential 1-form $A$ as $F=dA$. Thus, $Y(\mathcal{Q})$ function represents 
 the strength   of the non-minimal coupling between STPG and electromagnetism. The case of $Y' := d Y / d \mathcal{Q} =0$ or $Y=const.$ corresponds to minimal coupling.

In order to juxtapose the theory expressed in exterior algebra with the literature  using tensor formulation, we  write   $Q_{ab}=Q_{abc}e^c$ with $Q_{abc}$ $(0,3)$-type  non-metricity tensor 0-form; its first two indices are  symmetric $Q_{abc}=Q_{bac}$ by definition. Correspondingly, we calculate $\mathbb{Q}$ as
 \begin{align} \label{eq:lag-QQ2}
    \mathbb{Q} =& \big\{ (c_1+c_2+c_3) Q_{abc} Q^{abc} -c_2 Q_{abc}Q^{acb} -c_3 Q_{ab}{}^b Q^{ac}{}_c \nonumber \\
     & \qquad \qquad \qquad \qquad  + (c_4+c_5) Q^a{}_{ac} Q_b{}^{bc} -c_5 Q^a{}_{ab} Q^{bc}{}_c \big\} *1 .
 \end{align} 
We can prove that the STPG Lagrangian given in (\ref{eq:lag-QQ}) or (\ref{eq:lag-QQ2}) is corresponding to GR by setting  $c_i$'s as below \cite{Adak20062},
  \begin{align} \label{eq:gr-equiv-values-ci}
      c_1=  \frac{1}{2\kappa} , \quad c_2 =- \frac{1}{\kappa} , \quad c_3 =0 , \quad c_4 = - \frac{1}{2\kappa}, \quad c_5= \frac{1}{\kappa} ,
  \end{align}
where $\kappa$ is gravitational coupling constant in GR. If these $c_i$ values are inserted in the solutions of the model, the standard (or general relativistic) results can be obtained.

\subsection{Field Equations}
The variation of the Lagrangian (\ref{eq:lag-stpg}) can be calculated by the following  general formula given in \cite{Adak20062},
 \begin{align}
      \delta(\alpha \wedge *\beta) = \delta \alpha \wedge *\beta + \delta \beta \wedge *\alpha - \delta e^a \wedge \big[ (\iota_a\beta) \wedge *\alpha - (-1)^p \alpha \wedge (\iota_a *\beta) \big] 
 \end{align}
for any $\alpha$ and $\beta$  $p$-forms. Thus, by calculating  the variations  $\delta e^a$, $\delta \omega^a{}_b$, $\delta \lambda_a$, $\delta \rho^b{}_a$ and $\delta A$, we obtain  the following gravitational and electromagnetic field equations, respectively,
 \begin{subequations} \label{eq:equations-stpg}
 \begin{align}
     T^a &= 0 , \label{eq:zero-tors-stpg}\\
     R^a{}_b &= 0, \label{eq:zero-curv-stpg} \\
      \left(1 - \mathcal{F} Y' \right) \tau_a[Q] - Y\tau_a[F] - \mathcal{Q} \mathcal{F} Y' *e_a + \Lambda *e_a + D\lambda_a &= 0 , \label{eq:field-eqn-coframe-stpg} \\
     \left(1 - \mathcal{F} Y' \right) \Sigma^b{}_a[Q] +  e^b \wedge \lambda_a + D\rho^b{}_a &= 0, \label{eq:field-eqn-connec-stpg} \\
     d \left( Y*F \right) &=0 , \label{eq:maxwell}
 \end{align}
  \end{subequations}
where we have defined $\mathcal{F} = *\mathbb{F}$ or $\mathbb{F}=-\mathcal{F}*1$ and the energy-momentum 3-form $ \tau_a[F]$  is 
  \begin{align}
      \tau_a[F] = -(\iota_aF) \wedge *F + F \wedge (\iota_a*F)
  \end{align}
  for the Maxwell Lagrangian.
Additionally  the energy-momentum 3-form $\tau_a[Q]$ for the STPG model Lagrangian \ref{eq:lag-QQ} is found as the sum of each component
  \begin{align} 
 \tau_a[Q]=  \sum_{i=1}^5 c_i \overset{(i)}{\tau_a}[Q] \label{eq:tauQ-stpg}
 \end{align}
where the components are
  \begin{subequations} \label{eq:enr-mom-nonmet-sqred}
 \begin{align}
    \overset{(1)}{\tau_a}[Q] =& - \left( \iota_a  Q^{bc}  \right) \wedge * Q_{bc} - Q_{bc} \wedge \left( \iota_a* Q^{bc}  \right)  , \\
    \overset{(2)}{\tau_a}[Q] =& - 2Q_{ab} \wedge *\left( Q^{bc} \wedge e_c \right) - \left[ \iota_a\left( Q^{dc} \wedge e_c \right) \right] \wedge *\left( Q_{db} \wedge e^b \right) \nonumber \\
    &   \qquad \qquad + \left( Q_{db} \wedge e^b \right) \wedge \left[ \iota_a*\left( Q^{dc} \wedge e_c \right)\right]  , \\
    \overset{(3)}{\tau_a}[Q] =& - 2Q^{bc} \wedge *\left( Q_{ac} \wedge e_b \right) - \left[ \iota_a\left( Q^{dc} \wedge e^b \right) \right] \wedge *\left( Q_{db} \wedge e_c \right) \nonumber \\
    & \qquad \qquad + \left( Q_{db} \wedge e_c \right) \wedge \left[ \iota_a*\left( Q^{dc} \wedge e^b \right)\right]  , \\
   \overset{(4)}{\tau_a}[Q] =& - \left( \iota_a  Q \right) \wedge * Q - Q \wedge \left( \iota_a* Q  \right)  , \\
    \overset{(5)}{\tau_a}[Q] =& - Q \wedge *\left( Q_{ab} \wedge e^b \right) - Q_{ab} \wedge *\left( Q \wedge e^b \right) - \left[ \iota_a\left( Q_{bc} \wedge e^c \right) \right] \wedge *\left( Q \wedge e^b \right) \nonumber \\
    & \qquad \qquad + \left( Q \wedge e^b \right) \wedge \left[ \iota_a*\left( Q_{bc} \wedge e^c \right)\right]  .
 \end{align}
 \end{subequations}
The  angular (hyper) momentum 3-form  in (\ref{eq:field-eqn-connec-stpg} is found  as the following sum
  \begin{align}  
 \Sigma^b{}_a[Q]= \sum_{i=1}^5 c_i \overset{(i)}{\Sigma^b{}_a}[Q] \label{eq:SigmaQQ-stpg}
 \end{align}
where the each component of the sum is calculated 
  \begin{subequations} \label{eq:ang-mom-nonmet-sqred}
  \begin{align}
       \overset{(1)}{\Sigma^b{}_a}[Q] =& 2 *Q^b{}_a  ,\\
      \overset{(2)}{\Sigma^b{}_a}[Q] =&  e^b \wedge *\left( Q_{ac} \wedge e^c \right) +  e_a \wedge *\left( Q^{bc} \wedge e_c \right)  ,\\
      \overset{(3)}{\Sigma^b{}_a}[Q] =&  e^c \wedge *\left( Q_{ac} \wedge e^b \right) +  e_c \wedge *\left( Q^{bc} \wedge e_a \right), \\
    \overset{(4)}{\Sigma^b{}_a}[Q] =& 2 \delta^b_a *Q  ,\\
    \overset{(5)}{\Sigma^b{}_a}[Q] =& \delta^b_a e^c \wedge *\left( Q_{cd} \wedge e^d \right) + \frac{1}{2} \left[  e_a \wedge *\left( Q \wedge e^b \right) +  e^b \wedge *\left( Q \wedge e_a \right) \right]  .
  \end{align}
   \end{subequations}

We note that the  all angular momentum 3-forms are symmetric, $\overset{(i)}{\Sigma_{ab}}[Q] = \overset{(i)}{\Sigma_{ba}}[Q]$. We realise that the over-complication in the coframe equation (\ref{eq:field-eqn-coframe-stpg}) and the connection equation (\ref{eq:field-eqn-connec-stpg}) can be reduced considerably by the crucial assumption
  \begin{align}
      \mathcal{F} Y' = 1-k   \label{eq:simple-assumption}
  \end{align}
where $k$ is an arbitrary constant. Thus, the coframe and connection field equations turn out to be
  \begin{subequations} \label{eq:equations-stpg2}
 \begin{align}
           k \tau_a[Q] - Y\tau_a[F] + (1-k) \iota_a \mathbb{Q} + \Lambda *e_a + D\lambda_a &= 0 , \label{eq:field-eqn-coframe-stpg2} \\
     k \Sigma^b{}_a[Q] +  e^b \wedge \lambda_a + D\rho^b{}_a &= 0 . \label{eq:field-eqn-connec-stpg2}
 \end{align}
  \end{subequations}
Here  we have used $ \mathcal{Q} *e_a = -\iota_a \mathbb{Q}$.  By taking the covariant exterior derivative of (\ref{eq:field-eqn-connec-stpg2}),  we  reach 
 the following equation,
 \begin{align}
    k D\Sigma^b{}_a[Q]  - e^b \wedge D\lambda_a  = 0 
 \end{align}
where we have used  $De^b  =0$ and $D^2\rho^b{}_a = R^b{}_c \wedge \rho^c{}_a - R^c{}_a \wedge \rho^b{}_c =0$,  because  $R^a{}_b =0$ and $T^a=0$ in    the symmetric teleparallel geometry.  
Then, we hit $\iota_b$ and obtain $D\lambda_a$ explicitly
  \begin{align}
       D\lambda_a = k \iota_b D\Sigma^b{}_a[Q]   \label{eq:Dlambdaa-stpg}
  \end{align}
where we have used the identities, $\iota_a e^b=\delta^b_a$ and $e^a\wedge \iota_a \mathbb{T}  = p\mathbb{T} $ for a $p$-form $\mathbb{T} $. By substituting $D\lambda_a$  into (\ref{eq:field-eqn-coframe-stpg2}) we obtain the following form of our gravitational field equation  
  \begin{align}
       k \iota_b D \Sigma^b{}_a[Q] + k\tau_a[Q] - Y \tau_a[F] + (1-k) \iota_a \mathbb{Q} + \Lambda *e_a =0 . \label{eq:field-eqn-combined-stpg3}
  \end{align}
Consequently, the field equations we need to solve are (\ref{eq:zero-tors-stpg}), (\ref{eq:zero-curv-stpg}), (\ref{eq:maxwell}) and (\ref{eq:field-eqn-combined-stpg3}).  We can look for solutions for the following  settings of the parameter $k$ which corresponds to minimal and non-minimal couplings. 
  \begin{enumerate}
      \item Minimal Coupling $k=1$: The condition (\ref{eq:simple-assumption}) yields $Y'=0$ meaning $Y=const$.  Without loss of generality, we can set $Y=1$. Then, our theory becomes minimal coupling of STPG and electromagnetism.
      \item Non-minimal Coupling $k\neq 1$: In this case we look for solutions together with $\mathcal{F}Y'=1-k$. Thus, our theory contains  non-minimal coupling between STPG and electromagnetism.
  \end{enumerate}
Also,  it is important to note that the Maxwell equation (\ref{eq:maxwell}) can be re-expressed  as  
 \begin{align}
     \qquad d*G = 0.
 \end{align}
Here the non-minimal coupling of $Y(\mathcal{Q})$ to  the electromagnetic fields may be represented by the constitutive tensor $G=Y(\mathcal{Q})F$  in the specific medium  which have  the gravitational fields.

\subsection{Coincident gauge of symmetric teleparallel geometry}
The symmetric teleparallel geometry can be obtained from the constraints 
\begin{eqnarray}
    R^a{}_b = d\omega^a{}_b + \omega^a{}_c \wedge \omega^c{}_b &=&0,  \\
T^a=de^a + \omega^a{}_b\wedge e^b  &=&  0, \\
Q_{ab}=-\frac{1}{2}dg_{ab} + \frac{1}{2} (\omega_{ab} + \omega_{ba}) &\neq & 0,
\end{eqnarray} 
in  orthonormal frame. Because of the first constraint, $\omega_{ab}$ can not be solved analytically in terms of $e^a$. But, instead of the orthonormal frame, if we rewrite these equations in the coordinate frame   
 \begin{eqnarray}
     R^\alpha{}_\beta = d\omega^a{}_\beta + \omega^\alpha{}_\gamma \wedge \omega^\gamma{}_\beta &=&0, \ \\
T^\alpha= d(dx^\alpha) + \omega^\alpha{}_\beta\wedge dx^\beta &=& 0, \\
Q_{\alpha \beta}=-\frac{1}{2}dg_{\alpha \beta}  +\frac{1}{2} (\omega_{\alpha\beta} + \omega_{\beta \alpha}) &\neq & 0, 
 \end{eqnarray}
then we can see explicitly that $\omega^\alpha{}_\beta =0 $ satisfies the all three constraints with help of Poincare Lemma, $d(dx^\alpha)=d^2x^\alpha=0$ where $dx^\alpha$ is called the coordinate (holonomic) coframe. This setting is introduced as the coincident gauge \cite{Adak2006_1}, \cite{tomi-lavinia2017} which will be our guide in this paper in order to search the exact solutions. In the coordinate frame, metric is given by 
  $g = g_{\alpha \beta} dx^\alpha \otimes dx^\beta$. 
Now we pass to the orthonormal coframe by $e^a = h^a{}_\alpha (g) dx^\alpha$ where transformation elements, the so-called vierbein, form the general linear group, $h^a{}_\alpha(g) \in GL(4,\mathbb{R})$. So, the full connection  $\omega^a{}_b$ is  obtained from the relation $\omega^a{}_b(g) = 
h^a{}_\alpha(g) dh^\alpha{}_b(g)$, since  $ \omega^\alpha{}_\beta(g)=0$ in
the transformation rule  $\omega^a{}_b =  h^a{}_\alpha \omega^\alpha{}_\beta h^\beta{}_b +
h^a{}_\alpha dh^\alpha{}_b$.
As a result we can calculate the tensor-valued non-metricity 1-form via $Q_{ab}(g)=\frac{1}{2} \left( \omega_{ab}(g) + \omega_{ba}(g) \right) \neq 0$ in terms of metric functions.

\section{Spherically Symmetric Static Metric}

At this step, with the aim of getting exact solutions,  we make the following  metric ansatz in spherical coordinates $x^\mu=(t,r,\theta,\phi)$ as follows  
 \begin{align}
     g =-f^2(r)dt^2 + g^2(r)dr^2 + r^2 d\theta^2 + r^2 \sin^2\theta d\phi^2 \label{eq:metric}
 \end{align}
with the metric functions  $f(r)$ and $g(r)$, which will be determined from the field equations. 
Then,  we rewrite the metric in terms of the orthonormal coframe as 
  \begin{align}
      g=\eta_{ab} e^a \otimes e^b 
  \end{align} 
where  the orthonormal coframe is expressed in terms of metric functions and the coordinate coframe
 \begin{align} \label{eq:coframe-ansatz}
     e^0 = f(r) dt, \qquad e^1 = g(r) dr , \qquad e^2 = r  d \theta , \qquad e^3 = r \sin\theta d\phi .
 \end{align}  
By the transformation relations $e^a=h^a{}_\mu dx^\mu$ and $dx^\mu = h^\mu{}_a e^a$,  we can determine the vierbein and its inverse as 
\begin{align}
    h^a{}_\mu = 
     \begin{bmatrix}
      f & 0 & 0 & 0 \\
      0 & g & 0 & 0\\
      0 & 0 & r & 0 \\
      0 & 0 & 0 & r\sin\theta
     \end{bmatrix}
      \qquad \text{and} \qquad 
     h^\mu{}_a = 
     \begin{bmatrix}
      \frac{1}{f} & 0 & 0 & 0 \\
      0 & \frac{1}{g} & 0 & 0 \\
      0 & 0 & \frac{1}{r} & 0 \\
      0 & 0 & 0 & \frac{1}{r\sin\theta}
     \end{bmatrix}
 \end{align}
for the spherically symmetric metric. Correspondingly, we calculate  the full connection 1-form  via the vierbein  $\omega^a{}_b = h^a{}_\mu dh^\mu{}_b$ as below 
 \begin{align}
     \omega^a{}_b =
     \begin{bmatrix}
        - \frac{f'}{fg} e^1 & 0 & 0 & 0 \\
        0 & -\frac{g'}{g^2} e^1 & 0 & 0 \\
        0 & 0 & -\frac{1}{rg} e^1 & 0 \\
        0 & 0 & 0 & -\frac{1}{rg}e^1 - \frac{\cot\theta}{r} e^2
    \end{bmatrix}  \label{eq:full-connect-on1}
 \end{align}
and finally the non-metricity 1-form is obtained
 \begin{align}
     Q_{ab} = \frac{1}{2}(\omega_{ab} + \omega_{ba}) =  \begin{bmatrix}
         \frac{f'}{fg} e^1 & 0 & 0 & 0 \\
        0 & -\frac{g'}{g^2} e^1 & 0 & 0 \\
        0 & 0 & -\frac{1}{rg} e^1 & 0 \\
        0 & 0 & 0 & -\frac{1}{rg}e^1 - \frac{\cot\theta}{r} e^2
    \end{bmatrix} . \label{eq:Qab-conincident} 
    \end{align}
Also we saw that  $T^a=0$, $R^a{}_b=0$ constraints  are satisfied automatically. It is important to notice that the connection 1-form and the non-metricity 1-form are obtained in terms of the metric functions for the coincident gauge. 

Additionally, we   make the following electromagnetic potential 1-form  ansatz  in terms of $r$-dependent function $V(r)$ as 
  \begin{align}
      A = V(r) dt \label{eq:maxw-potential1}
  \end{align}
which leads to $F=  E(r) e^0 \wedge e^1$ where $E(r)=-V'(r)$ corresponds to the radial electric field.

\subsection{Minimal coupling: Reissner-Nordström-de Sitter solution}

In order to check the formalism, we consider the minimally coupled   case and we set $k=1$ that leads to $Y'(\mathcal{Q})=0$. Without loss of generality, we can set $Y=1$ which corresponds to the following model  
\begin{align} 
      L= \mathbb{Q} -  \mathbb{F} + \Lambda *1 + T^a \wedge \lambda_a + R^a{}_b \wedge \rho^b{}_a 
 \end{align}
that is symmetric teleparallel equivalent of Einstein-Maxwell theory obtaining by the replacement $\mathbb{Q} \rightarrow R *1$. We calculate the gravitational and electromagnetic field equations for the above metric and electromagnetic potential ansatz. Then we  set the coupling constants  $c_i$'s to the   GR-equivalent values given by  (\ref{eq:gr-equiv-values-ci}) in the results and   we see that   our minimal field equations are satisfied  by the Reissner-Nordström-de Sitter metric as expected 
 \begin{align} \label{eq:RN-dS-solution1}
     f(r)= \frac{1}{g(r)} = \sqrt{1- \frac{2m}{r} + \frac{\kappa q^2}{r^2} +  \frac{1}{3} \kappa \Lambda r^2}  \quad  \text{and} \quad E(r) = \frac{q}{r^2}
 \end{align}
where  $m$ and $q$ are integration constants corresponding to the mass and electric charge of the gravitating source, respectively. Thus the electrostatic potential function becomes $V(r) = {q}/{r} $.

\subsection{Non-minimal coupling and a class of solutions}

After crosschecking of our formalism in the minimal coupling via  $k = 1$, we now want to search new solutions for the parameter  $k\neq 1$ and leaving $Y(\mathcal{Q})$ arbitrary up to the constraint $dY/d\mathcal{Q}=(1-k)/\mathcal{F}$ for the spherically symmetric static metric ansatz (\ref{eq:metric}) and the Maxwell field ansatz (\ref{eq:maxw-potential1}). We calculate the components of the gravitational field equations  and find   quite long and complex differential equations. When we take $g={1}/{f}$, these equations get  a little simpler.   By collecting the components of the gravitational field equations with respect to the orders of $\sin\theta$ we realize that the $c_i$ constants must satisfy the constraint
  \begin{align}\label{constraint1}
      c_5= -(c_1 + c_2 + c_3 +c_4)  
  \end{align} 
due to the independence of these functions from angles. Furthermore when we rearrange the differential equations according to (\ref{constraint1}), we see that 
the only way to reduce the number of equations in the system of differential equations, which is very complicated and the number of equations is more than the number of unknowns, is to choose the $c_i$ constants  as follows: 
 \begin{align}
     c_2= -2c_1 , \qquad c_3=0 ,  \qquad c_4= -c_1 , \qquad c_5= 2c_1.
 \end{align}
That is, we can  set the coupling constants as 
 $c_1=1/2\kappa$, $c_2=-1/\kappa$, $c_3=0$, $c_4=-1/2\kappa$, $c_5=1/\kappa$ without loosing generality,
which are  GR-equivalent values given in (\ref{eq:gr-equiv-values-ci}).
This combination of the coupling  constants leads to
the following differential equations 
  \begin{subequations}
      \begin{eqnarray}
 {f^2}'' - \frac{2(f^2-1)}{r^2}   = \frac{4\kappa}{k}YE^2 , \label{firstdif}\\
 {f^2}''  + (8-\frac{4}{k})\frac{{f^2}'}{r}  +  (6 -\frac{4}{k}) \frac{f^2}{r^2} -\frac{2}{r^2} -\frac{4\kappa \Lambda}{k} = 0 , \label{seconddif}
   \end{eqnarray}
  \end{subequations}
where $E=-V'(r)$. The second differential equation (\ref{seconddif})
is non-homogeneous Cauchy-Euler differential equation and it has the following solution for the metric function $f(r)$
\begin{eqnarray}\label{solutionfunc}
     f^2(r) = \frac{k}{3k-2} - \frac{2m_0}{r} + \frac{\kappa q^2}{r^{(6k-4)/k}} + \frac{\kappa \Lambda r^2}{3(2k-1)} . 
\end{eqnarray}
 Since the Maxwell equation (\ref{eq:maxwell}) yields
  \begin{align}
      r^2 Y E = q \label{eq:sol-maxwell1}
  \end{align}
where $q$ is an integration constant which corresponds to the  charge of the electromagnetic source,
  we can find the non-minimal function $Y(r)$ by using (\ref{eq:sol-maxwell1})  in (\ref{firstdif})
\begin{eqnarray}\label{Yr}
    Y(r) =\frac{(3k-2)kq^2\kappa}{ (k^3-k^2)r^2 + \kappa q^2(30k^3 -59k^2 +38k-8)r^{-4+\frac{4}{k}}}  
\end{eqnarray}
and from  (\ref{eq:sol-maxwell1}) we find the electric field
\begin{eqnarray}
    E(r)=\frac{q}{Yr^2}= \frac{ (k^3-k^2)r^2 + \kappa q^2(30k^3 -59k^2 +38k-8)r^{-4+\frac{4}{k}}}{q(3k-2)r^2} .
\end{eqnarray}
We have checked that for $k=1$ the concerned functions become $Y(r)=1$ and $E(r) = {q}/{ r^2}$ as expected. We have also the constraint equation 
 \begin{eqnarray}\label{constraint2}
     \frac{dY}{d\mathcal{Q}}E^2 =1-k\;  .
 \end{eqnarray}
Here we remark that this constraint equation is not a linearly  independent from \ref{seconddif} and therefore it is satisfied  automatically. In order to prove that  we firstly  substitute (\ref{eq:sol-maxwell1}) in (\ref{firstdif}) and find 
\begin{eqnarray}\label{1/Y}
\frac{1}{Y} = \frac{k}{4\kappa  q^2} \left[  r^4{f^2}'' + 2r^2(f^2-1)  \right]  .
\end{eqnarray}
By taking the  differential  of (\ref{1/Y}), we obtain 
\begin{eqnarray}\label{dY/Y2ilk}
 \frac{dY}{Y^2} =  -\frac{k}{4\kappa q^2} \left[  4r^3{f^2}'' +r^4{f^2}''' -4r(f^2-1)  -2r^2{f^2}' \right] dr \ .
\end{eqnarray}
 On the other hand,  from (\ref{constraint2}) and   \ref{eq:sol-maxwell1}  we have
 \begin{eqnarray}
     \frac{dY}{Y^2} = \frac{1-k}{q^2}r^4d\mathcal{Q}\label{dY/Y2}\ .
 \end{eqnarray}
For the set of coupling constants (\ref{eq:gr-equiv-values-ci}), $\mathcal{Q}$ becomes
 \begin{eqnarray}
     \mathcal{Q} =-\frac{2c_1}{r^2} \left( r f^2 \right)' = -\frac{1}{\kappa} \left( \frac{{f^2}'}{r} + \frac{{f^2}}{r^2} \right).
 \end{eqnarray}
After  we take the differential of $\mathcal{Q}$ and substitute it in (\ref{dY/Y2}), we compare right hand sides of the equations (\ref{dY/Y2})   and (\ref{dY/Y2ilk}) and  we find 
\begin{eqnarray}
 {f^2}'''  + \left( 8-\frac{4}{k} \right)\frac{{f^2}''}{r}  - 2\frac{{f^2}'}{r^2} -2 \left( 6 -\frac{4}{k} \right) \frac{{f^2}}{r^3}  +  \frac{4}{r^3}  = 0\; .
\end{eqnarray}
Thus we see that the constraint equation (\ref{constraint2}) turns into  the derivative of the second differential equation (\ref{seconddif}) and  (\ref{solutionfunc}) is also its  solution.
Again for $k=1$ the metric function becomes the Reissner-Nordström-de Sitter metric
\begin{eqnarray}
    f^2(r) = 1- \frac{2m_0}{r } +\frac{q^2\kappa}{r^2}  +   \frac{1}{3} \kappa \Lambda r^2  .
\end{eqnarray}
If we want to re-express the non-minimal function $Y(r)$  given by (\ref{Yr}) in terms of $\mathcal{Q}$,  we can  integrate equation (\ref{eq:simple-assumption}) and find the non-minimal function in compact form
\begin{eqnarray}
    Y= \int \frac{1-k}{\mathcal{F}  }d\mathcal{Q} +C
\end{eqnarray}
where we remember that $\mathcal{Q} $  and $\mathcal{F}$   functions depend on the radial coordinate $r$ for the spherically symmetric metric,  $C$ is an integration constant and It can be fixed as $C=1$ to obtain the minimal coupling case for $Y=1$ or $k=1$
    \begin{eqnarray}
    Y=   1+  (1-k)\int  \frac{d\mathcal{Q}}{\mathcal{F} } .
\end{eqnarray}
Thus we obtain the non-minimal model in compact form as 
 \begin{eqnarray}
      L= \mathbb{Q} - \left( 1+  (1-k)\int  \frac{d\mathcal{Q}}{\mathcal{F} }\right) \mathbb{F} + \Lambda *1 + T^a \wedge \lambda_a + R^a{}_b \wedge \rho^b{}_a .
\end{eqnarray}
\section{Discussion}

In this paper, we have proposed a non-minimal model in the non-Riemannian geometry which is $ R^a{}_b =0$, $T^a= 0$,  $Q_{ab}\neq  0 $. The non-minimal model involves a non-minimally coupled electromagnetic field to the symmetric teleparallel gravity in  $Y(\mathcal{Q})F^2$ form.
After, we give a Lagrangian of our model, we derive field equations. Then, we look for spherically symmetric solutions only by making the electromagnetic potential and the metric ansatz by adhering the coincident gauge which allows us to determine the full (affine) connection in terms of metric functions. We find a class of solutions with  cosmological constant which are equivalent  to the solutions of the $Y(R)F^2$ model in Riemannian geometry \cite{Dereli20111,Sert12Plus,Sert13MPLA,Dereli20112}. In the minimal case with $k=1$, the solutions turn out to be the Reissner-Nordström-de Sitter solution as it should be. Accordingly, as a novel result, we have shown that we can transform the Riemannian non-minimal gravitoelectromagnetic model and the corresponding solutions in \cite{Dereli20111,Sert12Plus,Sert13MPLA,Dereli20112} to the symmetric teleparallel geometry which leads to a new intriguing research area to solve the  well known problems of gravity.







\end{document}